\documentclass[prb,sort&compress,numbers,amsmath,amssymb,superscriptaddress,twocolumn]{revtex4-1}
\usepackage{graphicx}
\usepackage{epsfig}
\usepackage[tight]{subfigure}
\usepackage{bm}
\usepackage{color}
\usepackage{tabularx}

\begin{document}
\widetext 

\title{First-Principles Description of Charge Transfer in Donor-Acceptor Compounds from 
Self-Consistent  Many-Body Perturbation Theory}

\author{Fabio Caruso}
\affiliation{Fritz-Haber-Institut der Max-Planck-Gesellschaft, Faradayweg 4-6, D-14195 Berlin, Germany}
\affiliation{Department of Materials, University of Oxford, Parks Road, Oxford OX1 3PH, United Kingdom}

\author{Viktor Atalla}
\affiliation{Fritz-Haber-Institut der Max-Planck-Gesellschaft, Faradayweg 4-6, D-14195 Berlin, Germany}

\author{Xinguo Ren}
\affiliation{Key Laboratory of Quantum Information, University of Science and Technology of China, Hefei, 230026, China}

\author{Angel Rubio}
\affiliation{Nano-Bio Spectroscopy group and ETSF Scientific Development Centre, Dpto.\ F\'isica de Materiales, Universidad del Pa\'is Vasco, CFM CSIC-UPV/EHU-MPC and DIPC, Av.\ Tolosa 72, E-20018 San Sebasti\'an, Spain}
\affiliation{Fritz-Haber-Institut der Max-Planck-Gesellschaft, Faradayweg 4-6, D-14195 Berlin, Germany}
\affiliation{European Theoretical Spectroscopy Facility}

\author{Matthias Scheffler}
\affiliation{Fritz-Haber-Institut der Max-Planck-Gesellschaft, Faradayweg 4-6, D-14195 Berlin, Germany}

\author{Patrick Rinke}
\affiliation{Fritz-Haber-Institut der Max-Planck-Gesellschaft, Faradayweg 4-6, D-14195 Berlin, Germany}
\date{\today}

\date{\today}
\begin{abstract}
We investigate charge transfer 
in prototypical molecular donor-acceptor compounds using hybrid
density functional theory (DFT) and the $GW$ approximation 
at the perturbative level 
($G_0W_0$) and at full self-consistency (sc-$GW$).
For the systems considered here, no charge transfer should be expected 
at large intermolecular separation according to 
photoemission experiment and accurate quantum-chemistry calculations.
The capability of hybrid exchange-correlation functionals of reproducing this
feature depends critically on the fraction of exact exchange $\alpha$, as
for small values of $\alpha$  spurious fractional charge transfer is observed between the 
donor and the acceptor. 
$G_0W_0$ based on hybrid DFT yields 
the correct alignment of the frontier orbitals for all values of $\alpha$. However, $G_0W_0$ has no capacity to alter the ground-state properties of the system, because of its perturbative nature.  
The electron density in donor-acceptor compounds thus remains incorrect for small $\alpha$ values. 
In sc-$GW$, where the Green's function is obtained from the iterative solution of  the Dyson equation,
the electron density is updated and reflects the correct description of 
the level alignment at the $GW$ level, demonstrating the importance of self-consistent
many-body approaches for the description of ground- 
and excited-state properties in donor-acceptor systems. 
\end{abstract}

\maketitle

\section{Introduction}

Donor-acceptor compounds have recently attracted considerable attention 
due to their application in the field 
of organic electronics \cite{ttf-tcnq-nature,doi:10.1021/cr050149z}. 
A description of donor-acceptor complexes from 
first principles is desirable 
to achieve an atomistic understanding of 
charge-transfer processes
and their impact on electronic properties.
 However, charge transfer remains a major challenge
 for presently available first principles 
 techniques \cite{doi:10.1021/ct1005517,1367-2630-15-12-123028}. 

In the weak-coupling limit (i.e., when the wave-function overlap between the
donor and the acceptor becomes negligible), the lowest 
charge-transfer energy ($E_{\rm CT}$) is
determined by the highest occupied molecular
orbital (HOMO) of the donor and the lowest 
unoccupied molecular orbital (LUMO) of the acceptor.
The HOMO and LUMO energies are equal to the negative of
the ionization potential (IP) and the electron affinity (EA).
In exact density functional theory (DFT) these values are given by the highest occupied 
Kohn-Sham (KS) levels of the $N$ and $N+1$ electron systems. 
For approximate DFT, the Slater-Janak transition states,\cite{PhysRevB.18.7165} i.e. the KS levels of the $N-\frac{1}{2}$
and $N+\frac{1}{2}$  electron systems, provide an accurate estimate of the HOMO and LUMO energies.  
At large separation between donor and acceptor, 
charge transfer may occur in the ground state if the HOMO of the donor lies energetically above the LUMO of the acceptor
or as a neutral charge transfer excitation otherwise.
Therefore, first principles methods that do not accurately capture orbital energies
of the $N-\frac{1}{2}$ and $N+\frac{1}{2}$ electron systems,
may provide a qualitatively wrong description of  
charge transfer and, subsequently, ground-state properties such as the charge density. 

An alternative to DFT for the description of the HOMO and LUMO (or IP and EA)
energies is many-body perturbation theory. Below we apply 
many-body Green's function theory to describe charge transfer. 
The single-particle Green's function provides a rigorous way to determine electronic excitations in molecules and solids and gives access to the total energy and therefore the ground-state properties of a system.  In this context, Hedin's $GW$ approximation \cite{Hedin1965} for the single-particle Green's function has become a well established framework for the calculation of IP and EA, 
also referred to as quasiparticle excitations  \cite{Aulbur/Jonsson/Wilkins:2000,Onida/Reining/Rubio,patrick2005}.  However, in perturbative $GW$ calculations  ($G_0W_0$) \cite{hybertsenlouie1986} only the quasiparticle energies are evaluated at the $GW$ level, whereas the ground-state density is left unchanged and remains at the unperturbed level, typically DFT. If spurious charge transfer has occurred at the DFT stage due to an inherent deficiency of  the chosen exchange-correlation (XC) functional, $G_0W_0$ cannot rectify this charge transfer, despite the fact that $G_0W_0$ may yield a  qualitatively correct HOMO-LUMO alignment in donor-acceptor systems. 

In this work, we demonstrate that the self-consistent $GW$ approach (sc-$GW$) -- 
in which the Green's function is obtained from the iterative solution 
of the Dyson equation -- provides a suitable first principles framework for 
the description of donor-acceptor systems. 
Compared to $G_0W_0$, the main advantages of the sc-$GW$ method
are the consistent description of ground and excited 
states and its independence 
of the initial reference ground state \cite{caruso/prb/2012}. 
For a set of prototypical donor-acceptor complexes,
we assess the performance of DFT  
hybrid exchange-correlation functionals, $G_0W_0$, and sc-$GW$ 
based on the following criteria: 
(i) accuracy of the quasiparticle spectrum and 
(ii) charge transfer. 
We show that sc-$GW$ yields a qualitatively correct HOMO-LUMO alignment. 
Moreover, it correctly predicts that the chosen donor-acceptor complexes do not exhibit any charge transfer at large donor-acceptor distances, as expected from reference experimental data and high-level quantum-chemical calculations.

The remainder of this Article is organized as follows.
In Sec.~\ref{sec:method}, we give an overview of the theoretical and computational methods employed in this work.  
Sec.~\ref{sec:DA} introduces the problem of charge transfer in donor-acceptor complexes 
in DFT and $GW$-based approaches.
The origin of charge transfer in DFT and in $GW$ is discussed in Sec.~\ref{sec:CTdft} and \ref{sec:CTgw}, respectively.
Finally, a summary and conclusions are reported in Sec.~\ref{sec:conc}.

\section{Computational Approach}\label{sec:method}

In this work, we apply the Perdew-Burke-Ernzerhof (PBE) hybrid family 
of XC functionals \cite{PBE,perdew:9982},
which expresses the XC energy as: 
\begin{align}\label{eq:Exc}
E_{\rm xc} = \alpha E_{\rm x}^{\rm EX} +(1-\alpha)  E_{\rm x}^{\rm PBE} + E_{\rm c}^{\rm PBE}\quad,
\end{align}
where $ E_{\rm x(c)}^{\rm PBE}$ is the PBE exchange (correlation) energy, 
$E_{\rm x}^{\rm EX}$ the exact exchange (EX) energy, and 
$\alpha\in[0,1]$ a real parameter. 
As an example, the PBE0 functional is obtained by
setting $\alpha=1/4$ in Eq.~(\ref{eq:Exc}). 

In $G_0W_0$, the quasiparticle energies 
$\epsilon_{n}^{{\rm QP}}$ are   
obtained from the first-order 
perturbative correction of the generalized Kohn-Sham (GKS) 
eigenvalues $\epsilon^0_{n}$: 
\begin{align}\label{eq:g0w0-qp}
\epsilon_{n}^{{\rm QP}}
= \epsilon^0_{n} + 
\left\langle \psi_n \right|
\hat\Sigma(\epsilon_{n}^{{\rm QP}}) - \hat v_{{\rm xc}}
\left| \psi_n \right\rangle
\quad,
\end{align}
where $\hat\Sigma$ is the $G_0W_0$ self-energy, $\hat v_{\rm xc}$ the XC 
potential of the preceding calculation, and $\psi_n$ the GKS  
orbitals. Here and below, spin indices have been omitted for simplicity.
In $G_0W_0$, the quasiparticle correction is applied only to the eigenvalues, 
whereas ground-state properties remain unaffected.
The perturbative inclusion of higher order terms in the self-energy
(such as second-order screened exchange (SOSEX), or vertex corrections 
\cite{paier,PhysRevLett.112.096401,sosex_SE})
are expected to improve the agreement of the computed excitation energies with experiment, but, similar to $G_0W_0$,
they would have no effect on the ground-state properties of the systems.
To incorporate the effect of the self-energy 
into the ground state, self-consistency is essential. 
In sc-$GW$ the Green's function $G$ is updated by solving the 
Dyson equation iteratively:
\begin{align} \label{eq:dyson}
&G(\epsilon) = G_0(\epsilon) + 
G_0(\epsilon)
\left[ \Sigma(\epsilon) + 
\Delta v_{\rm H}- v_{\rm xc}\right] G(\epsilon)\,\, ,
\end{align}
where $G_0$ is the Green's function of the DFT reference system, and $\Delta v_{\rm H}$
the difference of the $GW$ and DFT Hartree potentials.
At self-consistency, quasiparticle excitation energies are extracted directly from 
the spectral function $A(\omega)=1/\pi|{\rm Tr}[{\rm Im} G(\omega)]|$. 
Additionally,
from the self-consistent Green's function one may 
derive ground-state properties that are consistent with the $GW$ 
self-energy as, 
for instance, the electron density 
\begin{equation}\label{eq:density}
   n({\bf r})=-i G({\bf r},{\bf r},\tau=0^-) \quad,
\end{equation}
where $\tau$ denotes imaginary time 
(see Ref.~\onlinecite{PhysRevB.88.075105} for details).

\section{Donor-acceptor complexes}\label{sec:DA}

In the following, we consider prototypical donor-acceptor systems  obtained from a 
co-facial arrangement of donor and acceptor molecules.
In particular, we choose tetrathiafulvalene (TTF) 
as donor molecule and three different 
acceptors: tetracyanoethylene (TCNE), tetracyanoquinodimethane (TCNQ), and p-chloranil. 
All calculations are performed with 
the all-electron numeric atom-centered orbital code 
FHI-aims \cite{blum,1367-2630-14-5-053020,PhysRevB.88.075105}. 
The geometries of the individual molecules 
are obtained from a PBE geometry optimization performed with 
FHI-aims' Tier 2 basis set\cite{PhysRevB.88.075105}. For more details on the FHI-aims basis sets we refer the interested reader to Ref.~\onlinecite{blum,1367-2630-14-5-053020,Zhang/etal:2013}.
All geometries are listed in Appendix~\ref{app:geom}. 
For brevity, the following discussion is centered on the TTF-TCNE dimer. 

\begin{figure}
\includegraphics[width=0.48\textwidth]{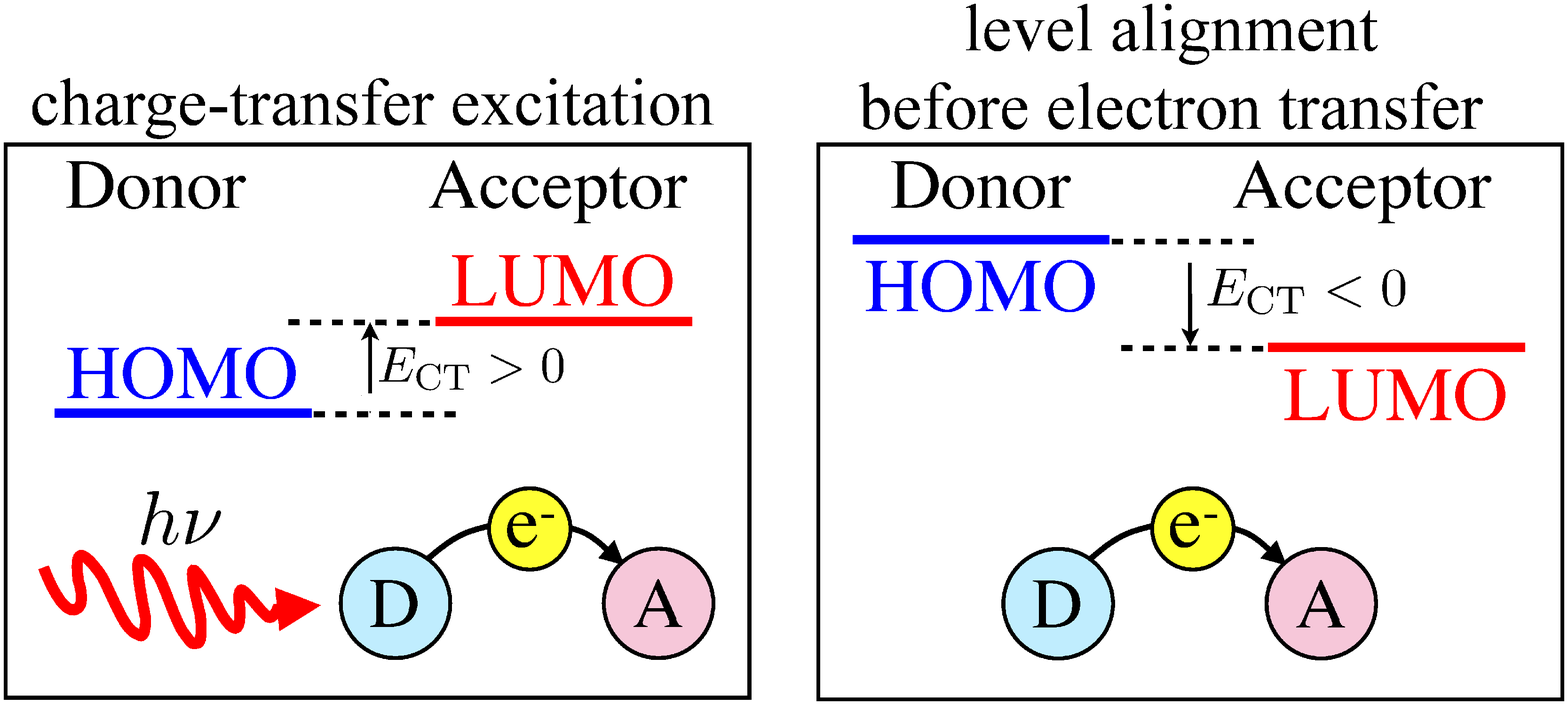}
\caption{Schematic representation of the level alignment in 
weakly-interacting donor-acceptor compounds. 
For $E_{\rm CT}>0$ (left), charge transfer occurs 
as an excitation (e.g., upon absorption 
of a photon with energy $h\nu\ge E_{\rm CT}$).
For negative values of the charge-transfer 
energy (right), the system is characterized 
by charge transfer in the ground state.
}
\label{fig:levelalignment}
\end{figure}

\subsection{TTF-TCNE dimer}

For small weakly-interacting molecules, the  charge-transfer energy at large intermolecular distances $R$ may  be approximated by: 
\begin{align}\label{eq:ect}
E_{\rm CT} = {\rm IP (donor)}  - {\rm  EA(acceptor)} - \frac{f^2}{R}\quad.
\end{align}
The last term is the Coulomb 
interaction arising from the transfer of $f$ electrons 
from the donor to the acceptor.

In the weak-coupling limit ($R\rightarrow \infty$) the Coulomb term can be neglected and  Eq.~\ref{eq:ect} reduces to
\begin{align}\label{eq-ectapp} 
E_{\rm CT}={\rm IP (donor) } - {\rm EA(acceptor) },
\end{align}
which now only depends on the relative energy position between the IP of 
the neutral donor and the EA of the neutral acceptor.
Charge transfer between the monomers occurs 
in the ground state whenever $E_{\rm CT}$ is 
negative [${\rm EA(acceptor)}>{\rm IP(donor)}$], whereas positive values of $E_{\rm CT}$ 
[${\rm EA(acceptor)}<{\rm IP(donor)}$]
indicate charge-transfer excitations. 
These two situations are schematically illustrated in 
Fig.~\ref{fig:levelalignment}.
In this limit  
-- according to experiment and coupled-cluster singles doubles with perturbative triples 
calculations (CCSD(T)) for the IP of TTF and  
the EAs of the donors considered here 
(see Table~\ref{tab:IE-EA}) --
no charge transfer should be expected at large 
intermolecular separation since $E_{\rm CT}>0$
for all donor-acceptor pairs.

\begin{figure} 
   \includegraphics[width=0.48\textwidth]{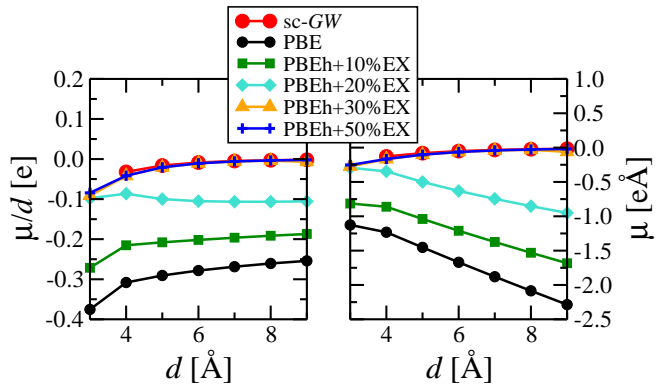} 
\caption{Left: the charge difference between TTF and TCNE is estimated from the 
ratio between the dipole moment and the distance  between the centers of two molecules.
Right: dipole moment of the TTF-TCNE dimer as function of the intermolecular distance.}
\label{fig:dipole}
\end{figure}

We first address the ground-state properties of the 
TTF-TCNE dimer.
For a quantitative assessment of the 
charge transfer between the donor and acceptor, 
we evaluated the dipole moment 
of the TTF-TCNE dimer for several 
values of the intermolecular distance
(right panel of Fig.~\ref{fig:dipole}).
Since charge transfer should not take place
in the weak-coupling limit, the component of the dipole 
moment parallel to the TTF-TCNE axis is expected to vanish. 
For PBE calculations and PBE-based 
hybrid functionals with $\alpha < 0.3$, however,
we observe 
a linear 
divergence of the dipole moment for increasing distance between the 
monomers. 
The diverging dipole at large intermolecular separation is 
a clear indication that charge is transferred from the donor to the acceptor.
The charge transfered between TTF and TCNE -- 
estimated from the ratio between the dipole moment 
and the intermolecular distance (left panel of Fig.~\ref{fig:dipole}) -- 
demonstrates that in PBE approximately one 
fourth of an electron is transferred from TTF to TCNE.
This picture is unaffected by the $G_0W_0$ quasiparticle 
correction of Eq.~\ref{eq:g0w0-qp}  because  $G_0W_0$ only corrects the DFT levels \textit{a posteriori}
but not the electron density which eventually determines the dipole moment.
On the other hand, the sc-$GW$ dipole moment 
-- derived from the sc-$GW$ density --
vanishes at large separation between the monomers and thus demonstrates
that charge transfer between the donor and the acceptor
is zero in the weak-coupling limit. On the DFT side, 
a charge-transfer-free description can be obtained from hybrid 
functionals that use a high $\alpha$-value ($>0.3$). 
For a more detailed discussion of charge transfer in DFT we refer to Section~\ref{sec:CTdft} and Ref.~\onlinecite{atalla:2014}. 

\begin{figure}
\includegraphics[width=0.43\textwidth]{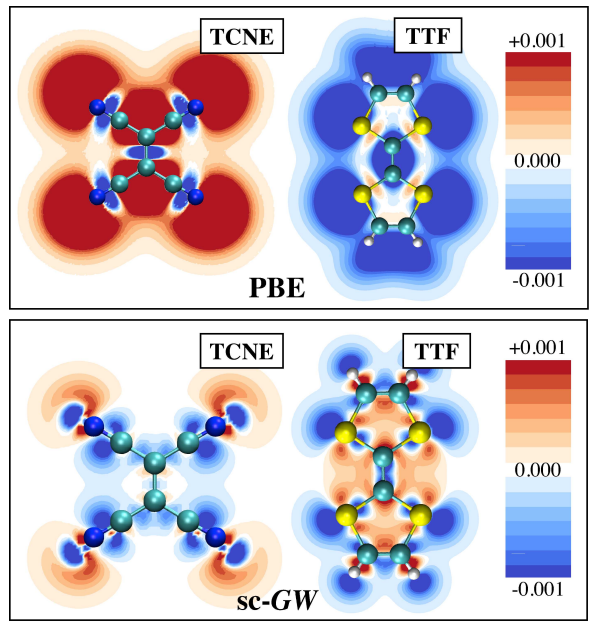}
\caption{Volume slices of the density difference 
between the TTF-TCNE dimer at 5 \AA\, distance and the 
monomers  in PBE (upper panel) and 
sc-$GW$ (lower panel).  
Units are \AA${}^{-3}$.}
\label{fig:densdiff}
\end{figure}

To illustrate the changes in the ground-state density induced by 
spurious fractional charge transfer, we report in
Fig.~\ref{fig:densdiff} volume slices of the difference between the 
TTF-TCNE density and the density of the isolated (neutral) monomers, evaluated from 
PBE (above) and sc-$GW$ (below) at a distance of $5$ \AA. 
The PBE density difference is mostly positive (red) on TCNE and negative (blue) on TTF. 
It therefore  manifests an accumulation of electron density on the 
acceptor accompanied by a charge depletion on the donor.
In sc-$GW$, however, the 
electron density is only slightly perturbed 
due to the weak interaction between the monomers. 
The density difference does not exhibit 
any charge transfer between TTF and TCNE.

\begin{figure}
\includegraphics[width=0.43\textwidth]{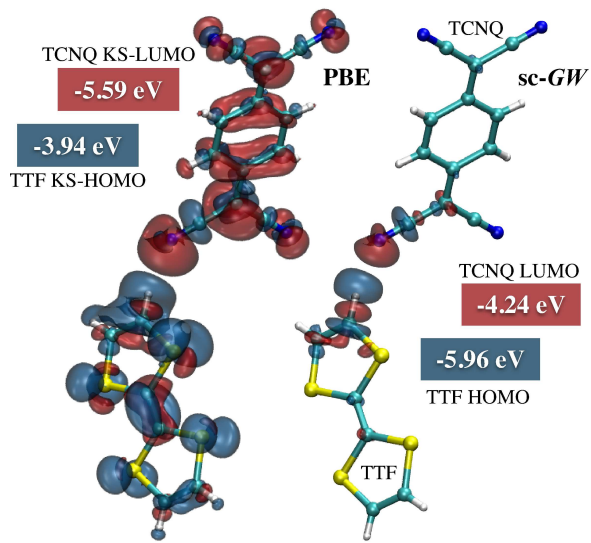}
\caption{Density difference between the TTF-TCNQ dimer and the monomers for PBE (left) 
and sc-$GW$ (right).  
Red (blue) isosurfaces indicate charge accumulation (depletion) resulting from the TTF-TCNQ level alignment. }
\label{fig:densdiff2} 
\end{figure}

\subsection{TTF-TCNQ dimer}

An additional example is provided by the TTF-TCNQ dimer shown in Fig.~\ref{fig:densdiff2}. 
The geometry of the dimer is taken from an interface between 
TTF and TCNQ crystals along the $[001]$ surface of TCNQ, 
in analogy with the work presented in Ref.~\onlinecite{atalla}. 
Figure~\ref{fig:densdiff2} reports isosurfaces
of the density difference between the dimer and the monomers. 
Like for the TTF-TCNE dimer, the highest occupied KS level of the isolated TTF molecule ($-3.94$ eV) 
incorrectly lies above the lowest unoccupied KS level of TCNQ ($-5.59$ eV) in PBE.
As a result, a fraction of an electron is transferred from TTF to TCNQ. 
In sc-$GW$ no charge transfer occurs, in analogy with the  TTF-TCNE dimer discussed before.
However, we observe a small charge rearrangement where 
the molecules are closest. This is most likely due to 
the Pauli principle, that requires that the molecular 
states of each molecule also have to be orthogonal 
to the states of the other molecule.

\section{Charge transfer in DFT}\label{sec:CTdft}

The spurious charge transfer that PBE and hybrid functionals with  a low fraction of exact exchange predict for donor-acceptor systems is related to  the {\it deviation from straight line error} (DSLE). In other words, the total energy does not exhibit a linear behavior for fractional electron numbers, as expected for the 
exact XC functional.   Janak's theorem\cite{PhysRevB.18.7165} establishes a relation between 
the Kohn-Sham eigenvalues and the total energy 
of a system with fractional electron number. 
For instance, for the HOMO level ($\epsilon^{\rm H}$) 
one has $\partial E(f)/\partial {f} = \epsilon^{\rm H}(f)$, where 
$E(f)$ denotes the total energy of a system in which the highest occupied KS level 
is occupied with $f$ electrons (with $N_0-1 < f \leq N_0$, 
$N_0$ being the integer number of electrons in the neutral system).
The corresponding eigenvalue of the molecule at integer occupation is then 
obtained in the limit $f\rightarrow N_0^-$. 
If the total energy is a linear function of $f$, 
the KS eigenvalue (i.e., the derivative of the total energy) is independent of $f$. 
Conversely, if the XC functional suffers from DSLE, the highest occupied and lowest unoccupied KS levels will exhibit an $f$ dependence, that will be stronger for larger deviations from linearity.  For XC functionals that produce a convex deviation from the straight line, the highest occupied KS level is too low in energy and the  lowest unoccupied KS level too high. For a concave deviation, such as in PBE, the  highest occupied KS level  is too high and the  lowest unoccupied KS level  too low. A concave deviation could thus result in spurious charge transfer \cite{atalla:2014}.

Hybrid functionals can be used to reduce or eliminate the DSLE. We achieve this by varying the fraction $\alpha$ of exact-exchange in the PBEh hybrid functional until the DSLE is minimized \cite{atalla:2014}. 
The resulting values ($\alpha^\dagger$) can be found in Tab.~\ref{tab:alpha}. 

\begin{table}[b]
\caption{Optimized $\alpha$ values of TTF, TCNE, TCNQ, 
and p-chloranil determined according to Refs.~\onlinecite{atalla} 
($\alpha^*$) and \onlinecite{PhysRevB.86.041110} ($\overline\alpha$).} 
\begin{tabular} {lcccc}
\hline\hline
& TTF \,\,\, & TCNE\,\,\, & TCNQ\,\,\, & p-chloranil\\ 
\hline
$\overline\alpha$ (Ref.~\onlinecite{PhysRevB.86.041110}) & 0.10 & 0.20 & 0.24 & 0.08 \\
$\alpha^*$ (Ref.~\onlinecite{atalla}) & 0.78 & 0.83 & 0.81 & 0.74 \\ 
$\alpha^\dagger$ (Ref.~\onlinecite{atalla:2014}) & 0.7 & -- & 0.7 & -- \\ 
\hline
\hline
\label{tab:alpha} 
\end{tabular}
\end{table}

%
\begin{figure}
\centering
\includegraphics[width=0.315\textwidth]{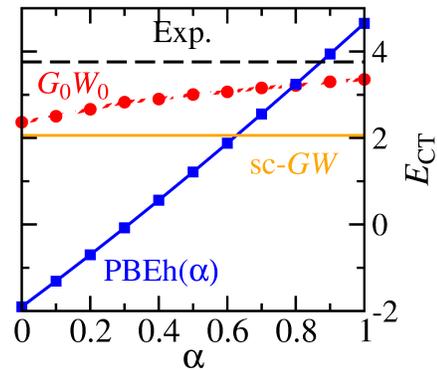}
\caption{
$G_0W_0$@PBEh($\alpha$) and 
and sc-$GW$ values for $E_{\rm CT}$ as a function of $\alpha$.
The PBEh($\alpha$) values are estimated from the difference between 
the lowest unoccupied and highest occupied KS levels.
The difference between the experimetal IP 
of TTF and EA of TCNE is included for comparison.} 
\label{fig:CTregion}
\end{figure}

\begin{table}[hb]
\caption{\label{tab:IE-EA} Comparison of experimental IPs and EAs with 
HOMO and LUMO energies of TTF and TCNE from different levels of theory. }
\begin{tabular} {lccr}
\hline\hline
    & HOMO$_{\rm TTF}$ & LUMO$_{\rm TCNE}$ & $E_{\rm CT}$ \\ 
\hline
PBE                                             & -3.94 &  -5.84 & -1.9   \\ 
PBE0                                            & -4.81 &  -5.20 & -0.39  \\ 
PBEh($\alpha^*$) \cite{atalla}                  & -6.82 &  -3.51 & 3.31   \\ 
PBEh($\overline\alpha$) \cite{PhysRevB.86.041110} & -4.28 &  -5.33 & -1.05  \\ 
$G_0W_0$@PBE                                 & -6.21 &  -3.85 & 2.36   \\
$G_0W_0$@PBE0                                & -6.42 &  -3.70 & 2.72   \\
$G_0W_0$@PBEh($\alpha^*$) \cite{atalla}      & -6.82 &  -3.54 & 3.28   \\
$G_0W_0$@PBEh($\overline\alpha$)                                      
\cite{PhysRevB.86.041110}                    & -6.28 &  -3.77 & 2.51  \\
sc-$GW$                                          & -5.96 &  -3.90 & 2.06  \\
\hline                                                                
Exp.  \cite{doi:10.1021/ja00165a007}           & -6.70 &   & (3.76)   \\
CCSD(T) \cite{Milian2004148}                            & &  -2.94  & (3.76)   \\
\hline
\hline
\end{tabular}
\end{table}

\section{Charge transfer in \textit{GW}}\label{sec:CTgw}

Figure~\ref{fig:CTregion} reports $E_{\rm CT}$ for $G_0W_0$ calculations based on PBEh hybrid calculations as a function of $\alpha$. For comparison, $E_{\rm CT}$  for PBEh itself is also shown. For low $\alpha$ $E_{\rm CT}$ is negative, which is consistent with spurious charge transfer found in DSLE functionals. $E_{\rm CT}$ increases linearly with $\alpha$ and approaches the experimental value for $\alpha$ values close to $\alpha^\dagger$.

In contrast, $G_0W_0$ calculations based on PBEh  always yield positive charge transfer energies
$E_{\rm CT}$ that are in quantitative agreement with the reference data  for all values of $\alpha$ (see, e.g., 
Table~\ref{tab:IE-EA} and Fig.~\ref{fig:CTregion}).
However, since $G_0W_0$ calculations only correct the DFT eigenvalues, the $G_0W_0$ approach cannot  repair the 
occurance of spurious charge transfer. To achieve a qualitatively correct description of both 
ground- and excited-state properties of donor-acceptor systems in $G_0W_0$, 
it is therefore essential to base $G_0W_0$ on hybrid functionals with large 
$\alpha$ values.

These results illustrate that the choice of 
$\alpha$ in Eq.~(\ref{eq:Exc}) is critical 
for the accuracy of ground-state properties. 
Similarly, it has been demonstrated that 
the accuracy achievable in the description of $G_0W_0$ 
quasiparticle excitation energies 
also exhibits a  strong  $\alpha$ dependence
\cite{PhysRevB.86.041110,PhysRevB.86.245127,atalla}.
To ameliorate this shortcoming, several strategies for a first principles 
determination of $\alpha$  based on the $G_0W_0$ approximation have been 
suggested recently.
In Ref.~\onlinecite{atalla}, 
some of us proposed to determine  $\alpha$ by minimizing the 
$G_0W_0$ quasiparticle correction of the highest occupied KS level 
(later referred to as $\alpha^*$-method). 
In their consistent starting point approach (CSP), 
K\"orzd\"orfer and Marom \cite{PhysRevB.86.041110} 
have suggested to determine $\alpha$ such that  
the hybrid functional eigenvalue spectrum is as close as possible to  
a rigid shift of the $G_0W_0$ spectrum for the valence states. Their $\alpha$ value is denoted $\overline\alpha$ in the following.
Alternatively, $\alpha$ may be chosen to enforce the  Koopmans' condition \cite{PhysRevLett.105.266802}  
by requiring that the highest occupied KS level energy agrees with the total-energy difference between the neutral and the singly-ionized system 
(i.e., the $\Delta$-self-consistent field, or $\Delta$-SCF, ionization energy).
However, due to the qualitative agreement of the $G_0W_0$ and $\Delta$-SCF
ionization energies, the latter method is expected to yield $\alpha$ values 
similar to the $\alpha^*$-method of Ref.~\onlinecite{atalla}.

For the donor and acceptors considered here, 
the  $\alpha^*$ and $\overline\alpha$ values 
are reported in Table~\ref{tab:alpha}.
For TTF and TCNE, the $\alpha^*$-method 
yields $\alpha^*=0.78$ and $\alpha^*=0.83$, respectively. 
The CSP approach gives $\overline\alpha=0.10$ 
for TTF and $\overline\alpha=0.20$ for TCNE. 
For both approaches, the tuned hybrid parameter 
for the TTF-TCNE dimer was obtained from an average
of the coefficients of the isolated molecules, i.e., 
$\alpha^*=0.8$ and $\overline\alpha=0.15$.
Due to the large fraction of EX, 
the $\alpha^*$-method 
produces the correct level alignment 
in donor-acceptor complexes. Moreover, the 
highest occupied (lowest unoccupied) KS levels
 of TTF (TCNE), and 
$E_{\rm CT}$ agree with 
the experimental reference values (Table~\ref{tab:IE-EA}). 
The CSP-approach, on the other hand, generally yields smaller $\alpha$ values 
(for the systems considered here $0.1<\overline\alpha<0.3$),
that do not
recover the correct level 
alignment between the donor and the acceptor.
Therefore, the associated hybrid functional ground-state is still 
characterized by spurious charge transfer.

sc-$GW$, like $G_0W_0$, always predicts positive charge-transfer energies
although the sc-$GW$ HOMO (LUMO) energy of TTF (TCNE) is slightly less accurate
than the corresponding quantity in $G_0W_0$ (see, e.g.,
Table~\ref{tab:IE-EA}). 
However, the Green's function obtained from the 
solution of the Dyson equation, Eq.~\ref{eq:dyson}, is  
independent
of the starting point and, therefore, of the EX parameter 
$\alpha$ (Fig.~\ref{fig:CTregion}). 
In addition, the electron density is also updated  through 
Eq.~\ref{eq:density}. Therefore, in contrast to $G_0W_0$, 
the ground-state properties are consistent with the correct 
level alignment, i.e., they are charge-transfer free.
We notice, however, that sc-$GW$ has a tendency to underestimate 
(overestimate) the IP (EA) for TTF-TCNE. The resulting $E_{\rm CT}$ is underestimated 
by approximately 2 eV compared to the reference value, 
showing that sc-$GW$ may still yield spurious ground-state charge transfer for 
donor-acceptor systems with small positive charge-transfer energies. 

\begin{figure}
\includegraphics[width=0.48\textwidth]{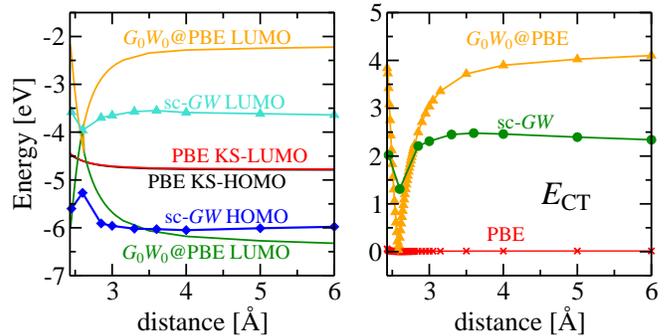}
\caption{Left: PBE, $G_0W_0$@PBE, and sc-$GW$ HOMO and LUMO levels for the TTF-TCNE dimer as 
a function of the distance between the monomers. Right: $E_{\rm CT}$ as a function of distance.}
\label{fig:HOMOLUMOdimer}
\end{figure}
Monitoring the frontier-orbital energies of the TTF-TCNE dimer 
as a function of distance between the monomers (Fig.~\ref{fig:HOMOLUMOdimer}, left) 
reveals an additional failure of the perturbative $G_0W_0$ approach. 
For all intermolecular separations, 
PBE yields degenerate highest occupied and lowest unoccupied KS levels. 
For distances larger than 3 \AA, $G_0W_0$ calculations break the HOMO-LUMO degeneracy of the PBE starting point and yield positive charge-transfer energies (Fig.~\ref{fig:HOMOLUMOdimer}, right).
However, for small distances, 
$G_0W_0$@PBE also yields vanishing charge-transfer energies and, therefore, 
might provide an overestimation of the charge transfer between donors and acceptors. 
In sc-$GW$, on the other hand, the separation between the HOMO and LUMO always remains finite and, correspondingly, $E_{\rm CT}>0$ at all distances.

\begin{figure}
\includegraphics[width=0.48\textwidth]{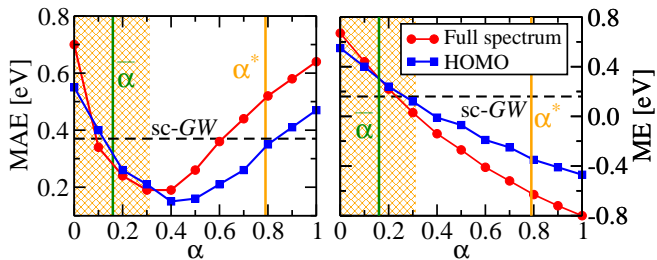}
\caption{Mean absolute error (MAE, left) and mean error (ME, right) 
relative to experiment 
\cite{ttf/exp,TCNE/TCNQ/spectrum,chloranil/pes,19842591} 
of the $G_0W_0$@PBEh($\alpha$) quasiparticle energies 
of TTF, TCNE, TCNQ, and p-chloranil as a function of $\alpha$ for the HOMO level (squares) and the full 
excitation spectrum (circles). 
The dashed lines refer to the sc-$GW$ errors for the 
full excitation spectrum.
The (average) optimally tuned parameters $\alpha^*$ and $\overline\alpha$ --  
determined according to Refs.~\onlinecite{atalla} 
and \onlinecite{PhysRevB.86.041110}, respectively -- are reported as vertical solid lines, 
whereas the shaded region indicates the range of $\alpha$ yielding 
spurious asymptotic charge transfer for TTF and TCNE.  }
\label{fig:MAEvsalpha}
\end{figure}
%

We next discuss the accuracy of the 
quasiparticle excitation spectrum 
of the donor and acceptor molecules.
For TTF, TCNE, TCNQ, and p-chloranil, the mean absolute 
error (MAE) and mean error (ME) of the $G_0W_0$@PBEh($\alpha$) 
quasiparticle energies relative to photo-emission experiment for the
HOMO level and for the valence excitation spectrum
are reported in Fig.~\ref{fig:MAEvsalpha} for 
$\alpha\in[0,1]$. The individual numbers are given in Tab.\,\ref{tab:MAEs} in the Appendix.
For the valence states, 
the MAE and ME refer to the first $N$ excitation energies
($N=10$ for TCNE and TCNQ, $N=6$ for TTF, $N=7$ for p-chloranil) 
for which  experimental data are available.
The best agreement with experiment -- with a MAE of about 0.2 eV -- 
is obtained with $\alpha=0.35$ for the HOMO, and $\alpha=0.4$ 
if all experimentally available excitation energies
are considered.
The sc-$GW$ ionization energies have a MAE of 0.4 eV, which 
-- being independent from the choice of $\alpha$ -- 
provides an unbiased assessment of the accuracy 
of the $GW$ approximation of the excitation spectrum of these systems.
For $\alpha > 0.5 $ the MAE is considerably larger for 
the valence spectrum than for the HOMO alone, 
indicating that large fractions of EX deteriorate 
the description of lower-lying excitations significantly, as demonstrated 
previously for benzene and the azabenzenes \cite{PhysRevB.86.245127}.
The increase of the MAE for large $\alpha$ (and similarly for small $\alpha$) 
values may be justified in terms of
the under-screening of the screened 
Coulomb interaction $W$ arising from the overstretching of the 
Kohn-Sham spectrum. 
The effect of over- and under-screening is clearly illustrated by the quasi-linear $\alpha$-dependence of the ME 
of $G_0W_0$ calculations 
(Fig.~\ref{fig:MAEvsalpha}, right), indicating that 
quasiparticle energies are underestimated (overestimated) for small (large) $\alpha$ values.

\section{Conclusions}\label{sec:conc}

In summary, we investigated the reliability of  
density functional approaches, 
$G_0W_0$, and sc-$GW$ in describing ground- and 
excited-state properties for a 
set of prototypical donor-acceptor compounds.
For donor-acceptor systems composed of TTF, TCNE, p-chloranil, and TCNQ, 
PBE-based hybrid functionals yield a spurious asymptotic charge transfer
for small values of the EX parameter $\alpha$, due to a misalignment 
of the frontier orbitals of the donor and the acceptor. 
The correct HOMO-LUMO alignment can 
be restored by resorting to hybrid XC functionals with large fractions of 
EX ($\alpha>0.3$). The accuracy of the full $G_0W_0$  excitation spectrum  reduces if high-$\alpha$
functionals were taken as starting point. 
Methods based on the $GW$ approximation provide a 
qualitatively correct description of the HOMO-LUMO alignment in all the
donor-acceptor systems considered here.
However, only sc-$GW$ describes the electron density correctly, 
since the ground-state is not updated in $G_0W_0$ calculations.
On the whole, sc-$GW$ is a promising scheme for the treatment
of donor-acceptor compounds, and more generally systems in which the
description of ground-state properties depends on the relative alignment of the frontier 
orbitals of different components (such as interfaces, molecules absorbed on surfaces, etc.).
Due to the cost of the sc-$GW$ calculations, however, 
future efforts should focus on 
the development of computationally affordable 
$GW$-based self-consistent approaches that 
will make larger systems tractable.

\acknowledgements
This work was supported by  the European Research Council Advanced Grant DYNamo (ERC-2010-AdG-267374), 
the European Commission within the FP7 CRONOS project (ID 280879), 
Spanish Grant (FIS2010-21282-C02-01), 
Ikerbasque and Grupos Consolidados UPV/EHU del Gobierno Vasco (IT578-13)

\appendix 

\section{Mean absolute errors of the ionization energies}

\begin{table*}[hbt]
\caption{\label{tab:MAEs} Mean absolute error (MAE) of the sc-$GW$ and
$G_0W_0$ quasi-particle energies energies based on different
starting points as compared to the first $N$
IPs ($N=10$ for TCNE and TCNQ, $N=6$ for TTF, $N=7$ for p-chloranil) experimentally available
from Refs.~\onlinecite{ttf/exp,TCNE/TCNQ/spectrum,chloranil/pes,19842591}.}
\begin{tabularx} 
           {\textwidth} {@{}%
          l@{\extracolsep{\fill}}%
          c@{\extracolsep{\fill}}%
          c@{\extracolsep{\fill}}%
          c@{\extracolsep{\fill}}%
          c@{\extracolsep{\fill}}%
          c@{\extracolsep{\fill}}}
\hline\hline
&TCNE   &TCNQ   &P-chloranil    &TTF    &Average  \\
\hline
sc-$GW$                  &  0.30 &  0.36         & 0.42  & 0.42  & 0.37   \\
$G_0W_0$@PBE             &  1.03 &  0.91         & 0.35  & 0.51  & 0.70   \\
$G_0W_0$@PBE0            &  0.35 &  0.38         & 0.38  & 0.16  & 0.32   \\
$G_0W_0$@HF              &  0.84 &  0.66         & 1.41  & 0.61  & 0.88   \\
$G_0W_0$@PBEh[$\alpha^*$]\cite{atalla}            &  0.58 &  0.42        & 1.06  & 0.39  & 0.61   \\
$G_0W_0$@PBEh[$\overline\alpha$] \cite{PhysRevB.86.041110}    &  0.48 &  0.40    & 0.25  & 0.37  & 0.37   \\
$G_0W_0$@PBEh($\alpha=0.2$)        &  0.48 &  0.36 & 0.30       & 0.20  & 0.24   \\
$G_0W_0$@PBEh($\alpha=0.4$)        &  0.05 &  0.23 & 0.61       & 0.09  & 0.19   \\
$G_0W_0$@PBEh($\alpha=0.6$)        &  0.30 &  0.41 & 0.90       & 0.24  & 0.36   \\
$G_0W_0$@PBEh($\alpha=0.8$)        &  0.56 &  0.62 & 1.16       & 0.36  & 0.52   \\
$G_0W_0$@PBEh($\alpha=1.0$)        &  0.74 &  0.79 & 1.37       & 0.46  & 0.64   \\
\hline\hline
\end{tabularx}
\end{table*}

In table \ref{tab:MAEs} we report the mean absolute errors (MAE) for the 
ionization energies of TTF, TCNE, p-chloranil, and TCNQ. 
The MAE refers to the first $N$ excitation energies
($N=10$ for TCNE and TCNQ, $N=6$ for TTF, $N=7$ for p-chloranil)
for which  experimental data are available \cite{Stafast1976855,TCNE/TCNQ/spectrum,doi:10.1021/ja00429a013,19842591}.

\section {Molecular geometries}
\label{app:geom}

Table\,\ref{tab:geo1}-\ref{tab:geo4} summarize the geometries of p-chloranine, TTF, TCNQ, and TCNE
optimized in the PBE approximation.

\begin{table}
\caption{\label{tab:geo1} PBE-optimized geometry of p-chloranine in cartesian coordinates and \AA.}
\begin{tabular} {r r r r}
\hline \hline
\textbf{p-chloranine} \\ \hline
 C         &  -1.86581       & 2.17291       &-0.04169  \\ 
 C         &  -0.93853       & 3.13223       & 0.04182  \\ 
 C         &   0.50931       & 2.80448       & 0.11635  \\ 
 C         &   0.88471       & 1.36593       & 0.09443  \\
 C         &  -0.04258       & 0.40661       & 0.01091  \\ 
 C         &  -1.49041       & 0.73437       &-0.06380  \\   
 O         &   1.36033       & 3.68495       & 0.19285  \\ 
 O         &  -2.34142       &-0.14609       &-0.14059  \\
 Cl        &  -3.56355       & 2.51305       &-0.13043  \\
 Cl        &  -1.33437       & 4.82019       & 0.07037  \\
 Cl        &   0.35325       &-1.28135       &-0.01743  \\
 Cl        &   2.58243       & 1.02579       & 0.18338  \\
\hline \hline
\end{tabular}
\end{table}

\begin{table}
\caption{\label{tab:geo2} PBE-optimized geometry of TTF in cartesian coordinates and \AA.}
\begin{tabular} {r r r r}
\hline
\hline \textbf{TTF} \\ \hline
 S     & -1.64052       &-1.48559      &  0.00163 \\ 
 S     & -1.64039       & 1.48575      &  0.00150 \\ 
 S     &  1.64051       & 1.48558      &  0.00164 \\ 
 S     &  1.64036       &-1.48575      &  0.00152 \\ 
 C     & -0.67944       & 0.00003      &  0.00113 \\ 
 C     &  0.67942       &-0.00003      &  0.00113 \\ 
 C     & -3.18226       &-0.67153      & -0.00158 \\ 
 C     & -3.18220       & 0.67182      & -0.00164 \\ 
 C     &  3.18225       & 0.67151      & -0.00157 \\ 
 C     &  3.18218       &-0.67184      & -0.00164 \\ 
 H     & -4.07491       &-1.29252      & -0.00288 \\ 
 H     & -4.07479       & 1.29289      & -0.00302 \\ 
 H     &  4.07490       & 1.29250      & -0.00288 \\ 
 H     &  4.07477       &-1.29291      & -0.00302 \\ 
\hline \hline
\end{tabular}
\end{table}

\begin{table}
\caption{\label{tab:geo3} PBE-optimized geometry of  TCNQ in cartesian coordinates and \AA.}
\begin{tabular} {r r r r}
\hline
\hline \textbf{TCNQ} \\ \hline
 C   &  0.00000      & 1.42000       &0.00000    \\    
 C   & -1.23502      & 0.68052       &0.00000    \\   
 C   &  1.23502      & 0.68052       &0.00000    \\   
 C   &  0.00000      &-1.42003       &0.00000    \\   
 C   & -1.23502      &-0.68053       &0.00000    \\   
 C   &  1.23502      &-0.68053       &0.00000    \\   
 C   &  0.00000      & 2.81981       &0.00000    \\   
 C   &  0.00000      &-2.81983       &0.00000    \\   
 C   & -1.20898      & 3.56632       &0.00000    \\   
 C   &  1.20896      & 3.56631       &0.00000    \\   
 C   & -1.20901      &-3.56628       &0.00000    \\   
 C   &  1.20899      &-3.56628       &0.00000    \\   
 N   & -2.20810      & 4.16937       &0.00000    \\   
 N   &  2.20811      & 4.16938       &0.00000    \\   
 N   & -2.20811      &-4.16938       &0.00000    \\   
 N   &  2.20812      &-4.16938       &0.00000    \\   
 H   & -2.17673      & 1.23005       &0.00000    \\   
 H   &  2.17673      & 1.23005       &0.00000    \\   
 H   & -2.17673      &-1.23005       &0.00000    \\   
 H   &  2.17673      &-1.23005       &0.00000    \\   
\hline \hline
\end{tabular}
\end{table}

\begin{table}
\caption{\label{tab:geo4} PBE-optimized geometry of  TCNE in cartesian coordinates and \AA.}
\begin{tabular} {r r r r}
\hline
\hline 
 \textbf{TCNE}  \\ \hline
 C    &  0.0000    	& 0.6901 	&0.0000 \\        
 C    &  0.0000 	&-0.6901 	&0.0000 \\        
 C    &  1.21464 	& 1.4330 	&0.0000 \\        
 C    &  1.21464 	&-1.4330 	&0.0000 \\        
 C    & -1.21464 	& 1.4330 	&0.0000 \\        
 C    & -1.21464 	&-1.4330 	&0.0000 \\        
 N    &  2.21911 	& 2.0231 	&0.0000 \\        
 N    &  2.21911 	&-2.0231 	&0.0000 \\        
 N    & -2.21911 	& 2.0231 	&0.0000 \\        
 N    & -2.21911        &-2.0231 	&0.0000 \\        
\hline \hline 
\end{tabular}
\end{table}


%

\end{document}